\documentclass{ifacconf}
\usepackage[utf8]{inputenc}
\usepackage{graphicx,natbib,amsmath,amssymb,booktabs,array,microtype,url,placeins}
\usepackage{xcolor}
\definecolor{linkblue}{rgb}{0,0.18,0.45}
\makeatletter
\let\ifac@ssect\@ssect
\makeatother
\usepackage[colorlinks=true,allcolors=linkblue]{hyperref}

\makeatletter
\AtBeginDocument{\let\@ssect\ifac@ssect}
\makeatother
\newcommand{\doi}[1]{doi:\discretionary{}{}{}\href{https://doi.org/#1}{\begingroup\urlstyle{rm}\nolinkurl{#1}\endgroup}}
\newcommand{\R}{\mathbb R}
\newcommand{\E}{\mathbb E}
\newcommand{\Cov}{\operatorname{Cov}}
\newcommand{\tr}{\operatorname{tr}}
\newcommand{\col}{\operatorname{col}}
\newcommand{\vct}{\operatorname{vec}}
\newcommand{\im}{\operatorname{im}}
\newcommand{\argmin}{\operatorname*{arg\,min}}
\newcommand{\dd}{\mathrm d}
\newcommand{\bu}{\mathbf u}
\newcommand{\hy}{\hat\theta}
\newcommand{\data}{\mathcal D}
\graphicspath{{figures/}}
\makeatletter
\newcommand{\embeddedfigurecaption}[2]{\begingroup
  \def\@captype{figure}\caption{#1}\label{#2}\endgroup}
\newcommand{\embeddedtablecaption}[2]{\begingroup
  \def\@captype{table}\caption{#1}\label{#2}\endgroup}
\makeatother
\begin{document}
\raggedbottom
\begin{frontmatter}
\title{\makebox[0pt][c]{\parbox{\textwidth}{\centering
The Prediction-to-Decision Gap:\\
First-Action Informativity in Data-Driven Predictive Control}}}
\author[KTH]{Aihui Liu}
\author[KTH]{Magnus Jansson}
\address[KTH]{Department of Information Science and Engineering,
KTH Royal Institute of Technology, Stockholm, Sweden
(e-mail: aihui@kth.se; janssonm@kth.se).}
\begin{abstract}
Model predictive control (MPC) applies only the first input of an optimized sequence before replanning. Do the available data determine this input? We define first-action informativity as agreement across data-compatible models at a fixed history and reference, with common MPC settings. For autoregressive models with exogenous inputs (ARX) and unconstrained quadratic MPC, we show that first-action agreement and prediction agreement under a fixed input sequence do not imply one another. We give structural conditions for action agreement without requiring unique model identification. When actions disagree, we quantify their mean-squared deviation from the fitted model’s input under a specific parameter distribution. We approximate this measure using local sensitivities and parameter covariance. Analytical and numerical examples illustrate how model uncertainty affects predictions and actions differently. The framework evaluates model uncertainty through its effect on the first action MPC is about to apply.
\end{abstract}
\end{frontmatter}

\section{Introduction}\label{sec:intro}

Model predictive control (MPC) optimizes an input sequence but applies only its first input before updating the history and replanning. This raises a specific data-sufficiency question: \emph{do all models compatible with the available data recommend the same input to apply now?} A dataset can answer this question even when it does not determine a unique model.

Prediction agreement under a supplied input sequence does not generally answer the same question. Models can predict identical outputs for that sequence while disagreeing about how changing the inputs changes those outputs. Their optimized inputs can then differ. Conversely, prediction differences can affect unactuated outputs or later planned inputs while leaving the first input unchanged. 

This question builds on control-oriented identification, where acceptable model errors depend on the intended application \citep{wahlberg2010input}. For MPC, \citet{ebadat2014application} use optimizer sensitivities to approximate application costs for identification-input design. Their cost averages squared output differences along trajectories. We build on this sensitivity approach to assess the current input across compatible models. The structural question is whether these inputs agree; sensitivity then helps quantify their remaining disagreement conditional on one dataset.

Existing work formalizes data sufficiency for different analysis,
prediction and control tasks. \citet{vanwaarde2020informativity}
develop an informativity framework based on the systems compatible
with the available data. For prediction, \citet{stevens2026prediction}
give sufficient conditions for unique output trajectories under
prescribed inputs and initialization in noiseless linear systems.
Here, each compatible model optimizes its own input sequence using
the same history, references, horizon and cost weights. The question is whether their first inputs agree.

The effect of model uncertainty depends on how model errors
influence the control task. \citet{wagenmaker2021task} combine
task curvature with statistical information to guide exploration.
\citet{hertneck2018learning} use input-error bounds to obtain
statistical closed-loop guarantees for a learned approximation
of an MPC policy. Companion work by \citet{liu2026firstactionloss}
relates the true-model cost penalty of a first input, with later
inputs reoptimized, to Lyapunov-type storage decrease. This motivates
assessing how identification uncertainty affects the current MPC action.

In this paper, we study ARX predictors and unconstrained quadratic MPC.
First, we define first-action informativity at a specified MPC decision
instance. For noiseless ARX data, we characterize when parameter changes
that the data cannot distinguish leave the first optimal input unchanged
across all compatible models. Analytical examples show that first-action
agreement and prediction agreement under a fixed input sequence do
not imply one another. An exact action-difference identity explains
this distinction.

Second, we quantify first-action uncertainty by the mean-squared
disagreement with the fitted model's input under a specified parameter
distribution. For noisy identification data, we use a fitted Gaussian
parameter distribution and derive local risk approximations using
action sensitivities and parameter covariance. Numerical examples
illustrate how model uncertainty affects predictions and first actions
differently across decision instances, and assess the approximation
accuracy.

Section~\ref{sec:setup} defines the model and MPC task.
Section~\ref{sec:structural} studies first-action agreement and
its relation to prediction agreement.
Section~\ref{sec:uncertainty} quantifies input disagreement,
and Section~\ref{sec:experiments} presents the numerical examples.

\section{Model, data and the control task}\label{sec:setup}\label{sec:arx}
\subsection{Identification data and ARX model}
Consider a strictly causal multiple-input multiple-output (MIMO) ARX model with $n_y$ outputs, $n_u$ inputs and lag orders $n_a\ge0$, $n_b\ge1$:
\begin{equation}\label{eq:arx}
 y_s=\sum_{\ell=1}^{n_a}\mathsf A_\ell y_{s-\ell}
     +\sum_{\ell=1}^{n_b}\mathsf B_\ell u_{s-\ell}+e_s.
\end{equation}
Here $y_s,e_s\in\R^{n_y}$, $u_s\in\R^{n_u}$, $\mathsf A_\ell\in\R^{n_y\times n_y}$ and $\mathsf B_\ell\in\R^{n_y\times n_u}$; $e_s$ is an innovation with zero conditional mean given the recorded past. Stack the unknown coefficients and the measured regressors as
\begin{align}
 \theta&=\col(\vct\mathsf A_1,\ldots,\vct\mathsf A_{n_a},
                  \vct\mathsf B_1,\ldots,\vct\mathsf B_{n_b}),\nonumber\\
 \varphi_s&=\col(y_{s-1},\ldots,y_{s-n_a},u_{s-1},\ldots,u_{s-n_b}),\nonumber\\
 \Psi_s&=\varphi_s^\top\otimes I_{n_y},\qquad y_s=\Psi_s\theta+e_s.\label{eq:regressor}
\end{align}
The operators $\col$, $\vct$ and $\otimes$ denote vertical stacking, column-wise vectorization and the Kronecker product. Set $d_h=n_an_y+n_bn_u$ and $p=n_yd_h$, so $\varphi_s\in\R^{d_h}$ and $\theta\in\R^p$. 

Let $\data$ contain $n$ input--output observations and the initial history needed for the first regressors. Stacking the equations gives
\begin{equation}\label{eq:data}
 \underbrace{\begin{bmatrix}y_1\\\vdots\\y_n\end{bmatrix}}_{y_\data}
 =\underbrace{\begin{bmatrix}\Psi_1\\\vdots\\\Psi_n\end{bmatrix}}_X\theta
 +\underbrace{\begin{bmatrix}e_1\\\vdots\\e_n\end{bmatrix}}_{e_\data}.
\end{equation}
Thus $X\in\R^{nn_y\times p}$ and $y_\data,e_\data\in\R^{nn_y}$. Let $\hy$ denote the vector of ARX coefficients estimated from the identification data. The corresponding fitted model serves as the nominal model. Rank deficiency can leave coefficient directions unidentified, while poor conditioning can leave substantial uncertainty under noise. We assess whether this uncertainty changes the applied input.

\subsection{Prediction from the current history}
At time $t$, the predictor is initialized with
\[
 h=\col(y_{t-1},\ldots,y_{t-n_a},u_{t-1},\ldots,u_{t-n_b}).
\]
For horizon $N\ge2$, let $\bu=\col(u_t,\ldots,u_{t+N-1})\in\R^{Nn_u}$ be a proposed future input sequence. Starting from $h$, repeatedly apply~\eqref{eq:arx} with future innovations set to zero. The resulting conditional mean prediction of $y_t,\ldots,y_{t+N-1}$ has the form
\begin{equation}\label{eq:prediction}
 \mu_\theta(h,\bu)=f_\theta(h)+G_\theta\bu,
\end{equation}
where $f_\theta\in\R^{Nn_y}$ and $G_\theta\in\R^{Nn_y\times Nn_u}$. The vector $f_\theta(h)$ is the prediction with zero future inputs; $G_\theta$ maps the planned inputs to outputs. Both can be nonlinear in the ARX coefficients; Appendix~\ref{app:derivatives} gives their construction.

Strict causality fixes the timing: $u_t$ can first affect $y_{t+1}$. Consequently, the first predicted output depends only on the history, and the last planned input does not affect outputs within this window. 

\subsection{MPC and the first input actually applied}\label{sec:mpc}
Let $r\in\R^{Nn_y}$ and $\bu_{\rm ref}\in\R^{Nn_u}$ be the output and input reference sequences. For symmetric weights $Q\succeq0$ and $R\succ0$ of the corresponding dimensions, consider the unconstrained objective
\begin{equation}\label{eq:objective}
 J_\theta(\bu)=\tfrac12\|f_\theta(h)+G_\theta\bu-r\|_Q^2
                  +\tfrac12\|\bu-\bu_{\rm ref}\|_R^2,
\end{equation}
where $\|v\|_W^2=v^\top Wv$. For a given model, the optimization gives
\begin{align}
 H_\theta&=G_\theta^\top QG_\theta+R,\nonumber\\
 k_\theta&=G_\theta^\top Q(r-f_\theta(h))+R\bu_{\rm ref},\nonumber\\
 \bu_\theta^*&=\argmin_{\bu}J_\theta(\bu)=H_\theta^{-1}k_\theta.\label{eq:mpc}
\end{align}
Since $H_\theta\succ0$, this solution is unique.

Collect the common history, reference trajectories, prediction horizon and cost weights into a \emph{decision instance}
\begin{equation}\label{eq:query}
 \xi=(h,r,N,Q,R,\bu_{\rm ref}).
\end{equation}
MPC applies only its first input block,
\begin{equation}\label{eq:firstaction}
 a_\theta(\xi)=E_0\bu_\theta^*,\qquad
 E_0=[I_{n_u}\ \ 0_{n_u\times(Nn_u-n_u)}],
\end{equation}
then updates the measured history and replans. The value $a_\theta(\xi)$ is the input prescribed by model $\theta$'s MPC policy at the current decision instance.

The identification data $\data$ need not determine a unique model. We now ask whether they nevertheless determine the first input to apply at a given decision instance $\xi$. We consider the input selected by each model compatible with the data, while keeping $\xi$ fixed. The supplied history $h$ initializes the predictions but is not used as additional identification data. This motivates the definition of first-action informativity in the next section.

\section{First-action informativity}\label{sec:structural}\label{sec:definition}

\subsection{Agreement across compatible models}
Let $\Theta_\data\subseteq\mathbb R^p$ denote a nonempty
set of candidate ARX coefficient vectors obtained from
the identification data $\data$. In the noiseless case,
these vectors satisfy $X\theta=y_\data$ within the assumed
parameter class.

\begin{defn}[First-action informativity]\label{def:informativity}
The data are first-action informative at $\xi$ if all
models in $\Theta_\data$ select the same first input:
\begin{equation}\label{eq:exactdefinition}
 a_\theta(\xi)=a_{\theta'}(\xi)
 \quad\text{for all }\theta,\theta'\in\Theta_\data.
\end{equation}
\end{defn}

\noindent\textit{Assumption (True-model inclusion).}
For the true-model interpretation of first-action informativity,
assume that the plant belongs to the chosen ARX model class
and that its coefficient vector $\theta_\star$ lies in $\Theta_\data$.

Under this assumption, first-action informativity implies that
the common input equals $a_{\theta_\star}(\xi)$.
If the fitted parameter $\hy$ also
belongs to $\Theta_\data$, its prescribed input therefore
agrees with that of true-model MPC.

\subsection{How model differences change the first action}\label{sec:sensitivity}
The next identity explains why identical output predictions under a fixed input sequence need not lead to the same first action. Prediction comparisons use the nominal optimal input sequence $\bu_{\hy}^*$ unless stated otherwise; each model optimizes its own input sequence when selecting its first action. The identity separates differences in predicted outputs from differences in their response to input changes. Keep $\xi$ fixed and define
\begin{align}
 \Delta a_\theta&=a_\theta-a_{\hy},\; \Delta G_\theta=G_\theta-G_{\hy},\nonumber\\
 \Delta\mu_\theta&=\mu_\theta(h,\bu_{\hy}^*)-\mu_{\hy}(h,\bu_{\hy}^*), \; e_{\rm tr}=\mu_{\hy}(h,\bu_{\hy}^*)-r.\label{eq:changes}
\end{align}
Here $e_{\rm tr}$ is the nominal tracking residual, not the innovation $e_s$. Evaluating both models' objective gradients at the nominal input sequence gives the difference
\begin{equation}\label{eq:gradientdifference}
 g_{\theta,\hy}=G_\theta^\top Q\Delta\mu_\theta
                    +\Delta G_\theta^\top Qe_{\rm tr}.
\end{equation}
\begin{prop}[Exact action-difference identity]\label{prop:finite}
For the predictor and objective in~\eqref{eq:prediction}--\eqref{eq:mpc},
\begin{equation}\label{eq:finiteaction}
 \Delta a_\theta=-E_0H_\theta^{-1}
 \bigl[G_\theta^\top Q\Delta\mu_\theta
       +\Delta G_\theta^\top Qe_{\rm tr}\bigr].
\end{equation}
\end{prop}
Models can disagree about the outputs predicted under the nominal input sequence. They can also disagree about how those outputs respond to a change in inputs. These differences produce the two terms in the objective-gradient change. Multiplication by $-H_\theta^{-1}$ gives the change in the optimized input sequence; $E_0$ selects the first input. In particular, prediction agreement at the nominal input sequence leaves
\[
 \Delta a_\theta=-E_0H_\theta^{-1}\Delta G_\theta^\top Qe_{\rm tr}.
\]
Thus identical predictions can coexist with different objective gradients and different first actions. Conversely, differences in later planned inputs can leave the first action unchanged. The proof subtracts the two optimality equations (Appendix~\ref{app:proofs}).

If predictions agree for \emph{every} input sequence, all models solve the same optimization problem and select the same input sequence. Also, if $Qe_{\rm tr}=0$, prediction agreement at the nominal input sequence already implies agreement of the entire optimal input sequence by~\eqref{eq:gradientdifference}.

\subsection{Prediction and first-action agreement do not imply each other}
\begin{exmp}[Same first action, different predictions]\label{ex:fir}\leavevmode\\
Consider a finite impulse response (FIR) model with three unknown input coefficients:
\[
 y_s=b_1u_{s-1}+b_2u_{s-2}+b_3u_{s-3}.
\]
Noiseless inputs $(u_{-3},u_{-2},u_{-1},u_0)=(0,0,1,0)$ and observations $(y_0,y_1)=(1,0)$ give $\operatorname{rank}X=2<3$: they identify $b_1=1$, $b_2=0$ but leave $b_3=\beta$ arbitrary. After applying $u_1=1$, consider $t=2$ before observing $y_2$. With $N=3$, $Q=R=I_3$, $\bu_{\rm ref}=0$ and $r=(0,1,1)^\top$,
\begin{equation}\label{eq:firexample}
 \mu_\beta(h,\bu)=\begin{pmatrix}1+\beta\\u_2\\u_3+\beta\end{pmatrix},
 \qquad \bu_\beta^*=\begin{pmatrix}1/2\\(1-\beta)/2\\0\end{pmatrix}.
\end{equation}
Every compatible model chooses $a_\beta=u_2^*=1/2$, although its predictions and second planned input depend on $\beta$. For other supplied histories and references, hold this family and the MPC settings fixed. No additional observations are assimilated. Then
\[
 a_\beta(\xi)=\frac{r_{t+1}-\beta u_{t-2}}{2}.
\]
First-action informativity holds exactly when $u_{t-2}=0$, for every reference.
\end{exmp}

\begin{exmp}[Same predictions, different first actions]\label{ex:converse}\leavevmode\\
Consider $y_s=bu_{s-1}$, a zero noiseless record, and zero initial history. Every $b\in\R$ is compatible. Let $N=2$, $Q=I_2$, $R=\rho I_2$, $\rho>0$, $r=(0,1)^\top$ and $\bu_{\rm ref}=0$. At $\hat b=0$, the nominal input sequence is zero, so every model predicts zero under that sequence. Reoptimization gives
\begin{equation}\label{eq:converseaction}
 a_b=\frac{b}{b^2+\rho},
\end{equation}
which varies with $b$. The models agree at the specified input sequence but disagree about the benefit of changing its first input.
\end{exmp}

Prediction agreement under one specified input sequence is therefore neither necessary nor sufficient for first-action agreement. These statements concern a specified decision instance, not long-run control performance.

\subsection{Unidentified directions: local and exact conditions}
Linearizing the action-difference identity gives a local comparison of prediction and action changes. For small parameter perturbations, define the two Jacobians
\begin{equation}\label{eq:jy}
 \begin{aligned}
 J_y&=\left.\frac{\partial\mu_\theta(h,\bu_{\hy}^*)}{\partial\theta}\right|_{\hy}
       \in\R^{Nn_y\times p},\\
 S&=\left.\frac{\partial a_\theta(\xi)}{\partial\theta}\right|_{\hy}
       \in\R^{n_u\times p}.
 \end{aligned}
\end{equation}
The planned inputs are held fixed in $J_y$; $S$ includes the change in the optimal input sequence. Differentiating the optimality condition gives
\begin{equation}\label{eq:sensitivity}
 S=-E_0H_{\hy}^{-1}(G_{\hy}^\top QJ_y+B_{\rm tr}),\qquad
 [B_{\rm tr}]_{:,i}=G_i^\top Qe_{\rm tr},
\end{equation}
with $G_i=\partial G_\theta/\partial\theta_i|_{\hy}$. Thus $\Delta\mu_\theta\approx J_y(\theta-\hy)$ and $\Delta a_\theta\approx S(\theta-\hy)$. Neither term in~\eqref{eq:sensitivity} universally dominates; they may reinforce or cancel. A prediction-sensitive direction can consequently have little effect on the first action.

For noiseless data and the unrestricted direct ARX coefficient class,
\begin{equation}\label{eq:affine}
 \Theta_\data=\{\theta:X\theta=y_\data\}=\hy+\im Z,
 \qquad \im Z=\ker X,
\end{equation}
where $\hy\in\Theta_\data$ and $Z\in\R^{p\times d}$ is a basis matrix, with $d=p-\operatorname{rank}X$. Moving along a column of $Z$ does not change the fitted outputs on the identification record. Proposition~\ref{prop:nullspaces} asks when these unidentified directions also leave predictions or actions unchanged, and separates a local action test from exact agreement over the full family.
\begin{prop}[Nullspace conditions]\label{prop:nullspaces}
On~\eqref{eq:affine}, with no stability or further parameter restrictions, the following conditions are necessary and sufficient for the stated properties:
\begin{align}
 J_y(\hy)Z=0&\quad\text{exact prediction agreement},\label{eq:predictionnull}\\
 S(\hy)Z=0&\quad\text{first-order action insensitivity},\label{eq:localnull}\\
 \frac{\partial a_\theta}{\partial\theta}Z=0 \; \forall\theta\in\Theta_\data
 &\quad\text{exact action agreement}.\label{eq:globalnull}
\end{align}
\end{prop}
The prediction condition~\eqref{eq:predictionnull} applies
to finite parameter changes. To see why, fix the history
and the nominal input sequence $\bu_{\hy}^*$. In the stacked
ARX equations, both the matrix $A_\theta$ and the
right-hand side are affine in $\theta$, and $A_\theta$
is unit lower triangular
(Appendix~\ref{app:derivatives}).
Subtracting the equations at $\hy+\delta$ and $\hy$
therefore gives the exact identity
\begin{equation}\label{eq:exactpredictionidentity}
 A_{\hy+\delta}(\mu_{\hy+\delta}-\mu_{\hy})
       =A_{\hy}J_y(\hy)\delta,
 \qquad \delta=Z\eta.
\end{equation}
Since both matrices are invertible, the prediction
difference is zero if and only if $J_y(\hy)\delta=0$.
Requiring this for every $\delta=Z\eta$ gives
$J_y(\hy)Z=0$, establishing exact prediction agreement
over the compatible family.

For the action map, Taylor expansion gives the local
condition~\eqref{eq:localnull}: $S(\hy)Z=0$ removes the
first-order action change at the nominal model.
Exact action agreement requires the directional
derivative to vanish throughout the affine family.
Integrating this derivative along segments of the
family establishes~\eqref{eq:globalnull}.

The single-point action condition is only local. The all-model condition is a structural characterization, not a finite computational certification procedure. Full column rank of $X$ suffices for both exact properties in the noiseless class, but Example~\ref{ex:fir} shows that it is unnecessary for first-action agreement.

Having characterized exact agreement across compatible models, we next quantify action uncertainty under a specified parameter distribution.

\section{Quantifying first-action uncertainty}\label{sec:uncertainty}
\subsection{Mean-squared action disagreement}
Let $p(\cdot\mid\data)$ be a specified parameter distribution, such as a posterior or a declared approximation to one, and let $\vartheta$ denote a random draw from it. Its support need not equal the compatible set $\Theta_\data$. Hold the decision instance $\xi$ and applied input $\hat a$ fixed. In the controller studied here, $\hat a=a_{\hy}(\xi)$.
\begin{defn}[First-action risk]\label{def:risk}
Assuming finite action second moments, define the mean-squared first-action disagreement with $\hat a$ under the specified distribution by
\begin{align}
 \mathcal R_u(\data,\xi;\hat a)
  &:=\E\!\left[\|a_\vartheta(\xi)-\hat a\|_2^2\mid\data\right]\nonumber\\
  &=\int\|a_\theta(\xi)-\hat a\|_2^2
 p(\dd\theta\mid\data).\label{eq:actionrisk}
\end{align}
\end{defn}
We call this mean-squared disagreement with the applied input the \emph{action risk}. Input- and output-channel scales are fixed in advance; each task's risk and tolerance use the corresponding scales.

For the unconstrained quadratic problem, Proposition~\ref{prop:finite} gives the exact specialization
\begin{equation}\label{eq:arxrisk}
 \mathcal R_u=\E[\|E_0H_\vartheta^{-1}g_{\vartheta,\hy}\|_2^2\mid\data].
\end{equation}
The local formulas below approximate this nonlinear target.

The distinction from action variance is useful. If $\bar a=\E[a_\vartheta(\xi)\mid\data]$, then
\begin{equation}\label{eq:biasvariance}
 \mathcal R_u=\tr\Cov(a_\vartheta(\xi)\mid\data)
                    +\|\bar a-\hat a\|_2^2.
\end{equation}
Even if $\hy$ is the mean parameter under the specified distribution, optimization is nonlinear in the model, so $a_{\hy}$ need not equal $\bar a$. Action variance alone can therefore miss part of the disagreement with the input actually applied.

For the same distribution and nominal input sequence, the prediction counterpart is
\begin{equation}\label{eq:predictionrisk}
 \mathcal R_y=\tfrac1N\E[\|\Delta\mu_\vartheta\|_2^2\mid\data].
\end{equation}
Only prediction is averaged over horizon steps. The two risks also have different physical units.
For specified root-mean-square (RMS) tolerances, compare
\begin{equation}\label{eq:tolerance}
 \mathcal R_u\le\varepsilon_u^2,\qquad
 \mathcal R_y\le\varepsilon_y^2.
\end{equation}
For an action map continuous in $\theta$, zero action risk
implies exact first-action informativity whenever the support
of $p(\cdot\mid\data)$ contains $\Theta_\data$.
Indeed, $\mathcal R_u=0$ means that
$a_\theta(\xi)=\hat a$ with probability one under $p$.
Continuity extends this equality to every model in the
support, and hence to every compatible model.

\subsection{A Gaussian parameter distribution}\label{sec:posterior}

For the numerical realization, we assume zero-mean Gaussian innovations and estimate their covariance from least-squares residuals (Appendix~\ref{app:moments}). The regressors contain only past inputs and outputs. We require $\widehat\Sigma_e\succ0$ and hold this estimate fixed in the Gaussian likelihood, so the resulting approximation ignores uncertainty in the covariance estimate. With the Gaussian prior $\theta\sim\mathcal N(\theta_0,P_0)$, where $P_0\succ0$, define
\begin{equation}\label{eq:posterior}
\begin{aligned}
\widehat\Omega &= I_n\otimes\widehat\Sigma_e,\;
\Lambda=X^\top\widehat\Omega^{-1}X,\; P=(P_0^{-1}+\Lambda)^{-1}, \\ \hy & =P(P_0^{-1}\theta_0+X^\top\widehat\Omega^{-1}y_\data),\;
\widetilde p(\theta\mid\data)  =\mathcal N(\theta;\hy,P).
\end{aligned}
\end{equation}
Here $\Lambda$ is the likelihood information matrix; Appendix~\ref{app:moments} gives the derivation. We evaluate the action risk in Definition~\ref{def:risk} using this fitted Gaussian distribution, taking $p=\widetilde p$. Since $\ker\Lambda=\ker X$, the prior keeps posterior uncertainty finite in directions that the data leave unidentified.

\subsection[Risk evaluation using linear and quadratic approximations]{\raggedright Risk evaluation using linear and quadratic approximations}\label{sec:local}\label{sec:approximations}
We first linearize the prediction and action maps to evaluate their risks, then include quadratic action terms for a further approximation. For the distribution in~\eqref{eq:posterior}, let $x=\vartheta-\hy$. The linear action and prediction changes $Sx$ and $J_yx$ from~\eqref{eq:jy} give
\begin{align}
 \mathcal R_u\approx\mathcal R_{u,1}
  &:=\tr(SPS^\top)=\sum_{i=1}^p\lambda_i\|Sv_i\|_2^2,\label{eq:linearrisk}\\
 \mathcal R_y\approx\mathcal R_{y,1}
  &:=\tfrac1N\tr(J_yPJ_y^\top)
     =\tfrac1N\sum_{i=1}^p\lambda_i\|J_yv_i\|_2^2,\label{eq:linearpred}
\end{align}
where $Pv_i=\lambda_i v_i$ and the $v_i$ are orthonormal. These expressions give the exact mean-squared disagreements for the linearized action and prediction maps, and approximate the corresponding nonlinear risks. Gaussianity is not needed for these linear moment identities. For a scalar input, $S$ is a row vector and $\mathcal R_{u,1}=SPS^\top$; for MIMO, the trace sums over input components.

In~\eqref{eq:linearrisk}, $\lambda_i$ measures parameter
uncertainty along $v_i$, while $Sv_i$ describes how the
first action responds to small parameter changes in that
direction. Their contribution to linear action risk is
$\lambda_i\|Sv_i\|_2^2$. Thus a highly uncertain parameter
direction can contribute little if the action is
insensitive to it.

The linear action change $Sx$ has zero mean because
$\E[x\mid\data]=0$. The nonlinear action change need not
have zero mean. A quadratic approximation can represent
this shift. Let $T_j$ be the Hessian of the $j$th action
component at $\hy$. We approximate the change from the nominal first action, $a_{\hy+x}(\xi)-a_{\hy}(\xi)$, by
\begin{equation}\label{eq:quadraticsurrogate}
 q(x)=Sx+\tfrac12\col(x^\top T_1x,\ldots,x^\top T_{n_u}x).
\end{equation}


For the centered Gaussian distribution in~\eqref{eq:posterior}, quadratic-form moments \citep{neuberger2025quadratic} and~\eqref{eq:biasvariance} give
\begin{equation}\label{eq:quadraticrisk}
 \begin{aligned}
 & \mathcal R_{u,2} :=\E[\|q(x)\|_2^2\mid\data]\\
 &=\tr(SPS^\top)+\tfrac12\sum_{j=1}^{n_u}\tr(T_jPT_jP) +\tfrac14\sum_{j=1}^{n_u}[\tr(T_jP)]^2.
 \end{aligned}
\end{equation}
The last term captures the squared mean shift predicted
by the quadratic action approximation. The formula is
exact for this quadratic action approximation, but is not a complete
second-order expansion of the nonlinear risk in $P$.
Section~\ref{sec:experiments} compares both approximations
with numerical integration; adding quadratic terms
need not improve accuracy.

\section{Numerical evaluation}\label{sec:experiments}
All examples here use the single-input single-output (SISO) ARX$(6,4)$ plant
\begin{align}
 a_\star&=(0,0.52,0.144,0.072,0.032,0.01)^\top,\nonumber\\
 b_\star&=(1.30,0.36,0.18,0.08)^\top,\quad
 e_s\overset{\rm iid}{\sim}\mathcal N(0,0.05^2).
 \label{eq:numericalplant}
\end{align}
MPC uses $N=30$, $Q=I_{30}$, $R=0.05I_{30}$ and zero input reference. For prediction comparisons, every model uses the nominal input sequence; for action comparisons, each model optimizes its own input sequence from the same history and reference.

\subsection{Prediction and first-action disagreement}\label{sec:exp1}

\begin{figure*}[tb]
\centering
\includegraphics[width=\textwidth]{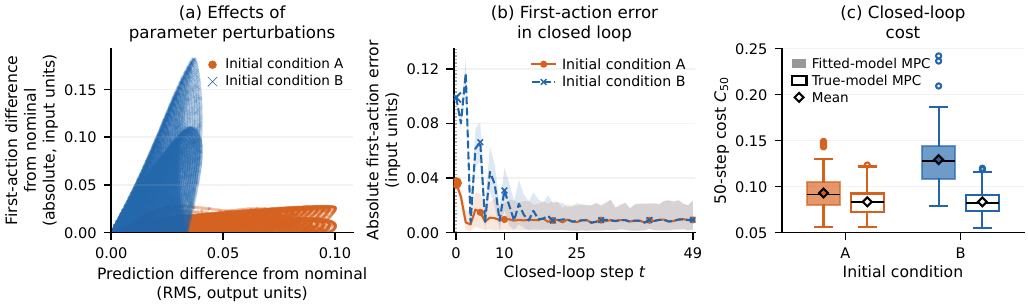}
\caption{(a) Prediction RMS and absolute first-action differences from the nominal model over the same 46,113 parameter perturbations at decision instances A and B. (b) First-action error from the true-model optimal input at the fitted controller's history and reference, over 200 noise realizations; curves show medians and bands show 10th--90th percentiles. (c) Cost $C_{50}=\tfrac12\sum_{t=0}^{49}[(y_t-1)^2+0.05u_t^2]$ for fixed fitted-model and true-model controllers over the same 200 noise realizations, each following its own trajectory.}
\label{fig:pair}
\end{figure*}

How does the same unresolved model uncertainty affect predictions and the first action at different decision instances?
We fit the nominal model $\hy$ by least squares using a long record of 12,000 input--output samples. The input is a sinusoid with frequency $1/32$ cycles per sample and RMS $0.5$. Despite the long record, the input is not persistently exciting of order four: its four lagged-input columns span only two dimensions. The regression matrix has rank $8<10$, leaving two parameter directions unidentified.

We compare two decision instances, A and B, that differ only in their histories:
\[
\begin{aligned}
 y_{\mathrm p}^{A}&=(0.943,1.605,1.127,3.015,-0.985,-0.934)^\top,\\
 u_{\mathrm p}^{A}&=(-0.012,-0.170,-0.420,0.756)^\top,\\
 y_{\mathrm p}^{B}&=(2.044,2.301,3.567,3.034,5.904,-4.498)^\top,\\
 u_{\mathrm p}^{B}&=(-0.636,-0.453,-0.437,-0.446)^\top.
\end{aligned}
\]
Both use the reference $r=\mathbf1_{30}$. We perturb the two unidentified parameter directions:
\begin{equation}\label{eq:lsfiber}
 \theta_\eta=\hat\theta+Z\eta,\quad \|\eta\|_2\le0.6,
 \qquad X\theta_\eta=X\hat\theta.
\end{equation}
Here $Z$ is an orthonormal basis of $\ker X$, so these perturbations preserve the identification fit. Subfigure~\ref{fig:pair}(a) compares predictions and first actions over the same parameter perturbations at A and B.

Figure~\ref{fig:pair}(a) shows a wider range of prediction differences but a narrower range of first-action differences at A than at B. Larger prediction differences therefore need not imply larger first-action differences at the current decision. In the closed-loop runs, B has larger initial action error and higher observed median fitted-controller cost than A; see Subfigures~\ref{fig:pair}(b,c).

\subsection{Which parameter directions matter?}\label{sec:directions}
\begin{figure}[tb]
\centering
\includegraphics[width=\columnwidth]{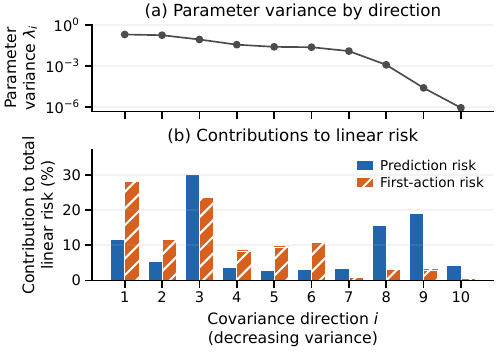}
\caption{Parameter uncertainty for one model fitted from 48 identification input--output samples. (a) Covariance eigenvalues in decreasing order. (b) Contributions are averaged over the 640 decisions and then normalized separately for each task. Each color's bars sum to $100\%$. Important directions differ between prediction and first action.}
\label{fig:directions}
\end{figure}

Which parameter directions matter for prediction and for the first action?
We now use the Gaussian parameter distribution in~\eqref{eq:posterior}, fitted to one short 48-sample record collected with the perturbed periodic input (Appendix~\ref{app:acquisition}). Figure~\ref{fig:directions} compares uncertainty along each covariance direction with its contribution to linear prediction and first-action risk.

We average each directional contribution in~\eqref{eq:linearrisk}--\eqref{eq:linearpred} over the 640 decisions. For each task, we divide these averages by their sum and multiply by 100 to obtain the percentages in Subfigure~\ref{fig:directions}(b).

Directions 8 and 9 contribute approximately $35\%$ of linear prediction risk but only $6\%$ of linear first-action risk, despite their small parameter variances. Parameter variance alone therefore does not determine which directions matter for a task.

\subsection{Accuracy and use of the risk estimates}\label{sec:accuracy}\label{sec:worked}
\begin{figure}[tb]
\centering
\includegraphics[width=\columnwidth]{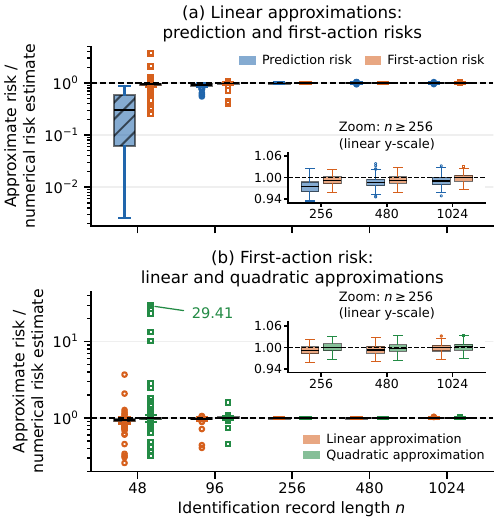}
\caption{Approximate risks divided by numerically estimated risks, with both averaged over the same 640 history--reference pairs. Each box summarizes all 128 fitted models; a ratio of one indicates agreement. (a) Linear prediction and first-action risk approximations. (b) Linear and quadratic first-action risk approximations.}
\label{fig:accuracy}
\end{figure}

How accurate are the first-order and second-order local risk approximations?
Figure~\ref{fig:accuracy} compares risks from linear and quadratic approximations with numerical risks under the fitted Gaussian distribution. At each $n\in\{48,96,256,480,1024\}$, 64 pairs of perturbed periodic and broadband records give 128 full-rank fits (Appendix~\ref{app:acquisition}). Both risks are averaged over the same 640 decision instances before taking their ratio. Agreement is close at the larger tested sample sizes.

At $n=48$, the estimated relative standard error exceeds $3\%$ for 104 of 128 numerical prediction-risk estimates and 9 of 128 first-action estimates. Striped boxes identify groups containing these estimates. Their limited numerical precision makes approximation accuracy harder to assess.

Both approximations describe behavior near the fitted model, so they can be inaccurate when the fitted parameter distribution is broad. Adding quadratic terms does not consistently improve the first-action risk estimates in these examples.

Table~\ref{tab:pointwise} compares the linear approximation with the numerically calculated first-action risk at individual decisions (Appendix~\ref{app:integration}). Among the included comparisons, the 95th-percentile relative error is below $6\%$ at the tested $n\ge256$. Most included individual decisions therefore also show close agreement at these record lengths.

\begin{table}[b]
\centering
\caption{Errors of the linear first-action risk approximation relative to the numerically calculated risk. Statistics use only the included comparisons.}
\label{tab:pointwise}
\begin{tabular}{@{}rrrr@{}}\toprule
 & Comparisons & \multicolumn{2}{c}{Relative error (\%)}\\
\cmidrule(lr){3-4}
$n$ & included (\%) & Median & 95th percentile\\\midrule
48   & 89.7 & 5.98 & 42.42\\
96   & 83.0 & 2.96 & 9.41\\
256  & 82.1 & 2.00 & 5.51\\
480  & 80.9 & 1.90 & 5.34\\
1024 & 80.6 & 1.84 & 5.28\\\bottomrule
\end{tabular}
\end{table}

Table~\ref{tab:decision} checks the tolerance criterion~\eqref{eq:tolerance} at one decision, using a perturbed periodic-input fit at $n=256$, zero history, reference $r_k=\sin(2\pi k/32)$, and RMS tolerances $\varepsilon_y=0.005$ and $\varepsilon_u=0.01$.

\begin{table}[tb]
\centering
\caption{Risk estimates divided by squared tolerances at one decision. Values $\le1$ meet the corresponding tolerance; bracketed ranges show only numerical uncertainty.}
\label{tab:decision}
\begin{tabular}{@{}lrr@{}}\toprule
Estimate & Prediction & First action\\\midrule
Linear & $1.155$ & $0.811$\\
Quadratic & --- & $0.817$\\
Numerical band & $[1.171,1.215]$ & $[0.813,0.891]$\\\midrule
\shortstack[l]{Estimates meet\\tolerance?} & No & Yes\\\bottomrule
\end{tabular}
\end{table}

Figure~\ref{fig:accuracy}(a) generally shows closer agreement between linear and numerical estimates for first-action risk than for prediction risk. The first-action map has only $n_u$ outputs, and cancellation of model effects may reduce nonlinear terms omitted by the approximation.

\section{Conclusion}\label{sec:conclusion}
First-action informativity asks whether all models compatible with the data choose the same first optimal input for a given history and reference under common MPC settings. First-action risk measures mean-squared deviation from the nominal action under a declared parameter distribution. For unconstrained quadratic MPC with ARX predictors, we show that agreement on predictions under a fixed input sequence and agreement on the first action do not imply one another. This assessment concerns the specified decision; it does not by itself certify task completion or closed-loop performance.

The definitions extend beyond ARX predictors. For other differentiable predictors, local approximations also require the optimal first action to vary sufficiently smoothly with model parameters. The numerical examples illustrate how the current history and task sensitivities change the relevance of the same model uncertainty. 

Future work includes constrained MPC with active-set changes and connecting the present risk to storage-decrease conditions \citep{liu2026firstactionloss}. Directions that affect the first action could also guide data collection and exploration as histories, references and model estimates change.

\appendix
\section{Proofs and additional calculations}
\subsection{ARX prediction and derivatives}\label{app:derivatives}
Stacking the zero-innovation ARX equations separates predicted-output, planned-input and fixed-history contributions into $\Phi_{y,\theta}$, $\Phi_{u,\theta}$ and $b_\theta(h)$, respectively. For $k,j=0,\ldots,N-1$, let block $(k,j)$ of $\Phi_{y,\theta}$ and $\Phi_{u,\theta}$ be $\mathsf A_{k-j}$ and $\mathsf B_{k-j}$ for $j<k$ when the lag is present, and zero otherwise. The lifted recursion is
\begin{align}
 [b_\theta(h)]_k&=\sum_{\ell=k+1}^{n_a}\mathsf A_\ell y_{t+k-\ell}
              +\sum_{\ell=k+1}^{n_b}\mathsf B_\ell u_{t+k-\ell},\nonumber\\
 A_\theta&=I-\Phi_{y,\theta},\quad
 A_\theta\mu_\theta=b_\theta+\Phi_{u,\theta}\bu,\nonumber\\
 f_\theta&=A_\theta^{-1}b_\theta,\qquad
 G_\theta=A_\theta^{-1}\Phi_{u,\theta}.\label{eq:lifted}
\end{align}
$A_\theta$, $\Phi_{u,\theta}$ and $b_\theta(h)$ are affine in $\theta$. Since $\Phi_{y,\theta}^N=0$, $A_\theta^{-1}=\sum_{k=0}^{N-1}\Phi_{y,\theta}^k$ exists for every finite $\theta$. With $R\succ0$, predictions and the optimizer are smooth without requiring model stability.

Subscripts $i$ and $ij$ denote first and second parameter derivatives at $\hy$ with $\xi$ fixed. With $F_i=\partial_i\Phi_y$ and $K_i=\partial_i\Phi_u$,
\begin{align}
 A_{\hy}f_i&=\partial_i b+F_i f_{\hy},&
 A_{\hy}G_i&=K_i+F_iG_{\hy},\nonumber\\
 A_{\hy}f_{ij}&=F_if_j+F_jf_i,&
 A_{\hy}G_{ij}&=F_iG_j+F_jG_i.\label{eq:predderivatives}
\end{align}
With $\bu_{\hy}^*$ fixed, $[J_y]_{:,i}=f_i+G_i\bu_{\hy}^*$. For the reoptimized sequence, its parameter derivatives $\bu_i$ and $\bu_{ij}$ follow by differentiating $H_\theta\bu_\theta^*=k_\theta$:
\begin{align}
 \bu_i&=H_{\hy}^{-1}(k_i-H_i\bu_{\hy}^*),\nonumber\\
 \bu_{ij}&=H_{\hy}^{-1}(k_{ij}-H_{ij}\bu_{\hy}^*-H_i\bu_j-H_j\bu_i),\nonumber\\
 [T_\ell]_{ij}&=[E_0\bu_{ij}]_\ell,\qquad \ell=1,\ldots,n_u.\label{eq:actionsecond}
\end{align}
Their first blocks give $S$ and $T_\ell$.

\subsection{Action-difference identity}\label{app:proofs}\label{app:structural}
\textit{Proposition~\ref{prop:finite}.} Subtract the nominal zero gradient from $\nabla_{\bu}J_\theta(\bu_{\hy}^*)$ to obtain $H_\theta\bu_{\hy}^*-k_\theta=g_{\theta,\hy}$. Subtract $H_\theta\bu_\theta^*=k_\theta$ and select the first block to prove~\eqref{eq:finiteaction}. Differentiating the optimality condition proves~\eqref{eq:sensitivity}.

\subsection{Posterior moments and interpretation}\label{app:moments}\label{app:quadratic}
Let $\Phi$ have rows $\varphi_s^\top$ and $r_X=\operatorname{rank}\Phi$ denote its rank for one output equation. The least-squares residuals are
\[
 \theta_{\rm LS}=X^\dagger y_\data,\qquad
 \hat e_s=y_s-\Psi_s\theta_{\rm LS}.
\]
For $n>r_X$, use the residual covariance estimate
\[
 \widehat\Sigma_e=\frac{1}{n-r_X}\sum_{s=1}^{n}\hat e_s\hat e_s^\top.
\]
Require $\widehat\Sigma_e\succ0$. For a scalar output, the estimate is the residual sum of squares divided by $n-\operatorname{rank}X$. This conventional degrees-of-freedom correction is not claimed exactly unbiased for stochastic ARX regressors.

Under Section~\ref{sec:posterior}'s assumptions, the sequential likelihood at covariance $\Sigma$ is proportional in $\theta$ to
\[
 \exp\{-\tfrac12(y_\data-X\theta)^\top
            (I_n\otimes\Sigma)^{-1}(y_\data-X\theta)\}.
\]
With $\Sigma=\widehat\Sigma_e$ fixed, completing the square with the Gaussian prior gives~\eqref{eq:posterior}. This plug-in construction also applies under causal feedback.

The action moments are finite: optimality against $\bu_{\rm ref}$ implies
\[
 \lambda_{\min}(R)\|\bu_\theta^*-\bu_{\rm ref}\|_2^2
 \le\|f_\theta+G_\theta\bu_{\rm ref}-r\|_Q^2.
\]
The right-hand side is polynomial in Gaussian parameters and has finite expectation. This permits unstable model draws in the finite-horizon uncertainty calculation.

For centered Gaussian $x$, odd moments eliminate the linear--quadratic cross terms in $\E\|q(x)\|^2$. The identities
\begin{align*}
 \E[x^\top T_jx\mid\data]&=\tr(T_jP),\\
 \E[(x^\top T_jx)^2\mid\data]&=2\tr(T_jPT_jP)+[\tr(T_jP)]^2
\end{align*}
give~\eqref{eq:quadraticrisk} by summing the component squares.

\section{Additional details for the main experiments}\label{app:numerics}
This appendix specifies the data and common evaluation decisions for Sections~\ref{sec:directions}--\ref{sec:accuracy}, and explains how numerical precision affects the comparisons.

\subsection{Data collection and evaluation}\label{app:acquisition}
We vary record length and identification input, but keep the evaluation decisions fixed. Before normalization, the two inputs are
\begin{align*}
 v_s^{\rm per}&=0.5+\sin(2\pi s/32+\phi)+0.05w_s,\\
 v_s^{\rm broad}&=w_s,\; w_s\overset{\rm iid}{\sim}\mathcal N(0,1),\; \phi\sim\mathrm{Unif}[0,2\pi).
\end{align*}
At each record length, 64 independent pairs contain one perturbed-periodic and one broadband record, giving 128 fits. Paired records share $w_s$ and plant innovations; sample sizes use separate acquisitions. Each record starts from zero history with 64 warm-up steps followed by $n$ fitting samples. Each input segment is normalized separately to RMS $0.5$. Fits use the prior $\mathcal N(0,100I_{10})$ and residual variance estimate $\hat\sigma_e^2=\|y_\data-XX^\dagger y_\data\|_2^2/(n-10)$.

The 640 common decision instances use history--reference pairs from two independent 320-step true-model MPC trajectories starting from zero history, with $r_t=\sin(2\pi t/32)$ and preview $r_{t,k}=r_{t+k}$. Fitted models and parameter distributions remain fixed during evaluation and subsequent closed-loop runs.

\subsection{Precision of the numerical calculations}\label{app:integration}
We check local approximations against numerical risks under the fitted Gaussian distribution. The precision target for the numerical calculation is an estimated standard error at most $3\%$ of each risk averaged over 640 decisions. Figure~\ref{fig:accuracy} includes all fits, even those missing this target.

For Table~\ref{tab:pointwise}, we take the absolute difference between the linear first-action risk $\mathcal R_{u,1}$ and the numerically calculated risk $\widehat{\mathcal R}_u$, divide by $\widehat{\mathcal R}_u$, and express the result as a percentage: $100|\mathcal R_{u,1}/\widehat{\mathcal R}_u-1|$. Only comparisons for which both averaged and individual-decision numerical action-risk estimates pass the $3\%$ precision check are included. The Comparisons included column reports the percentage used to calculate these error statistics.

Table~\ref{tab:decision}'s bracketed ranges are the numerical risk estimate $\pm1.96$ estimated standard errors, divided by the squared tolerance. These ranges show only numerical uncertainty.

\section{Additional comparison of data-collection inputs}\label{app:comparisons}
We now compare perturbed periodic and broadband identification inputs. Does the input with lower prediction error also give smaller first-action errors and lower closed-loop cost? Both inputs have RMS $0.5$. We use all 64 record pairs at $n=48$ from Appendix~\ref{app:acquisition}; this record length was chosen after comparing results across lengths. Errors use the same 640 decisions; costs follow each controller's own trajectories. Prediction MSE uses each fit's nominal inputs for both models over 30 outputs; first actions are reoptimized.

\begin{figure}[htb]
\centering
\includegraphics[width=\columnwidth]{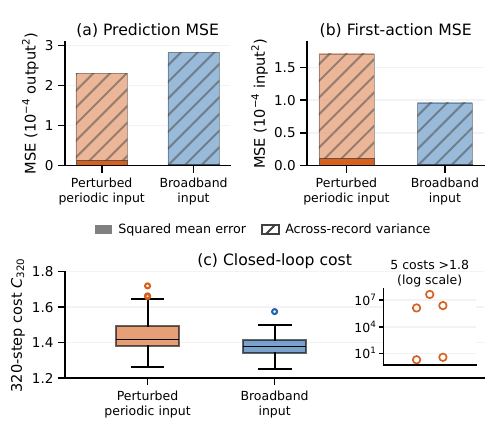}
\caption{Errors and costs for 64 fits per input, each based on $n=48$ identification samples. (a,b) MSE relative to the true model over the same 640 history--reference pairs: squared mean error (solid) and across-record variance (striped). (c) $C_{320}=\sum_{t=0}^{319}[(y_t-r_t)^2+0.05u_t^2]$, averaged over two closed-loop runs per fit from zero history with the sinusoidal reference. All costs enter the box statistics; the inset shows the five costs above $1.8$.}
\label{fig:inputcomparison}
\end{figure}

In this dataset, perturbed periodic input produces lower observed prediction MSE, higher first-action MSE and the five largest closed-loop costs. This motivates including first-action accuracy in identification-input design.

This selected example illustrates that prediction error, first-action error and closed-loop cost can favor different identification inputs. It does not establish that either input is consistently better.

{\raggedright\bibliography{references}}

@inproceedings{wahlberg2010input,
  author = {Wahlberg, Bo and Hjalmarsson, H{\aa}kan and Annergren, Mariette},
  title = {{On Optimal Input Design in System Identification for Control}},
  year = {2010},
  booktitle = {Proceedings of the 49th IEEE Conference on Decision and Control},
  pages = {5548--5553},
  publisher = {IEEE},
  doi = {10.1109/CDC.2010.5717863},
}

@inproceedings{ebadat2014application,
  author = {Ebadat, Afrooz and Annergren, Mariette and Larsson, Christian A. and Rojas, Cristian R. and Wahlberg, Bo and Hjalmarsson, H{\aa}kan and Molander, Mats and Sj{\"o}berg, Johan},
  title = {{Application set approximation in optimal input design for model predictive control}},
  year = {2014},
  booktitle = {Proceedings of the European Control Conference},
  pages = {744--749},
  publisher = {IEEE},
  doi = {10.1109/ECC.2014.6862496},
}

@article{vanwaarde2020informativity,
  author = {van Waarde, Henk J. and Eising, Jaap and Trentelman, Harry L. and Camlibel, M. Kanat},
  title = {{Data Informativity: A New Perspective on Data-Driven Analysis and Control}},
  year = {2020},
  journal = {IEEE Transactions on Automatic Control},
  volume = {65},
  number = {11},
  pages = {4753--4768},
  doi = {10.1109/TAC.2020.2966717},
}

@article{stevens2026prediction,
  author  = {Stevens, Joel and Coulson, Jeremy},
  title   = {{Informativity for Data-Driven Prediction}},
  journal = {IEEE Control Systems Letters},
  year    = {2026},
  volume  = {10},
  pages   = {1699--1704},
  doi     = {10.1109/LCSYS.2026.3708575},
}

@article{hertneck2018learning,
  author = {Hertneck, Michael and K{\"o}hler, Johannes and Trimpe, Sebastian and Allg{\"o}wer, Frank},
  title = {{Learning an Approximate Model Predictive Controller with Guarantees}},
  year = {2018},
  journal = {IEEE Control Systems Letters},
  volume = {2},
  number = {3},
  pages = {543--548},
  doi = {10.1109/LCSYS.2018.2843682},
}

@inproceedings{wagenmaker2021task,
  author = {Wagenmaker, Andrew J. and Simchowitz, Max and Jamieson, Kevin},
  title = {{Task-Optimal Exploration in Linear Dynamical Systems}},
  year = {2021},
  booktitle = {Proceedings of the 38th International Conference on Machine Learning},
  volume = {139},
  pages = {10641--10652},
  series = {Proceedings of Machine Learning Research},
  publisher = {PMLR},
  doi = {10.48550/arXiv.2102.05214},
  eprint = {2102.05214}
}

@article{neuberger2025quadratic,
  author = {Neuberger, J. Nicholas and Alexanderian, Alen and van Bloemen Waanders, Bart},
  title = {{Goal Oriented Optimal Design of Infinite-Dimensional Bayesian Inverse Problems using Quadratic Approximations}},
  year = {2025},
  journal = {Journal of Scientific Computing},
  volume = {105},
  pages = {55},
  doi = {10.1007/s10915-025-03073-y},
}

@unpublished{liu2026firstactionloss,
  author = {Liu, Aihui and Jansson, Magnus},
  title = {{First-Action Loss and Lyapunov-Type Decrease in Data-Driven MPC under Parameter Uncertainty}},
  year = {2026},
  note = {To be submitted},
}
\end{document}